\documentclass[twocolumn,showpacs,amsmath,amssymb,superscriptaddress, prd, aps]{revtex4-2}

\usepackage{xcolor}
\usepackage{enumerate}
\usepackage{graphicx}
\usepackage{dcolumn}
\usepackage{bm}
\usepackage[version=3]{mhchem} 
\usepackage{amsmath}
\usepackage{esvect}

\usepackage{array}
\usepackage{graphicx}
\usepackage{graphicx}
\usepackage{dcolumn}
\usepackage{bm}
\usepackage{color}
\usepackage{pifont}
\usepackage{natbib}
\usepackage{hyperref}
\hypersetup{
colorlinks = true,
urlcolor   = blue,
linkcolor  = blue,
citecolor  = blue
}
\usepackage{nicefrac}

\usepackage{color}

\begin{document}

\title{On local conservation of information content in Schwarzschild black holes}

\author{Godwill Mbiti Kanyolo}
\email{gmkanyolo@mail.uec.jp}
\affiliation{The University of Electro-Communications, Department of Engineering Science,\\
1-5-1 Chofugaoka, Chofu, Tokyo 182-8585, Japan}

\author{Titus Masese}
\email{titus.masese@aist.go.jp}
\affiliation{Research Institute of Electrochemical Energy (RIECEN), National Institute of Advanced Industrial Science and Technology (AIST), 1-8-31 Midorigaoka, Ikeda, Osaka 563-8577, Japan}
\affiliation{AIST-Kyoto University Chemical Energy Materials Open Innovation Laboratory (ChEM-OIL), Yoshidahonmachi, Sakyo-ku, Kyoto-shi 606-8501, Japan}


\begin{abstract}
The central equations in classical general relativity are the Einstein Field Equations, which accurately describe not only the generation of pseudo-Riemannian curvature by matter and radiation manifesting as gravitational effects, but more importantly mass-energy dynamics, evolution and distribution on the space-time manifold. Herein, we introduce a geometric phase in general relativity corresponding to Schwarzschild black hole information content. This quantity appropriately satisfies a local conservation law subject to minimal coupling, with other desirable properties such as the quantization of the black hole horizon in units of Planck area. The local conservation law is imposed by field equations, which not only contain the trace of Einstein Field Equations, but also a complex-valued function with properties analogous to the quantum-mechanical wave function. Such success attests to the utility of the proposed field equations in capturing key aspects of quantum gravity theories.
\end{abstract}

\maketitle

\textbf{\textit{Introduction}} -- Quantum information within the Schrodinger picture of quantum mechanics is locally conserved by virtue of the probability density, which is normalized to unity, satisfying the continuity equation.\cite{mcmahon2008quantum} This guarantees that the sum of probabilities of all possible configurations of the quantum system, even as the wave function evolves, is always preserved at unity. In the Heisenberg picture, this requires the existence of a unitary operator responsible for the quantum evolution of the wave function.\cite{braunstein2007quantum} In particular, the ultimate test of information conservation within a quantum system is the no-hiding theorem\cite{braunstein2007quantum, samal2011experimental}, which explicitly requires that any apparent loss of the quantum system's information content due to processes such as decoherence be a unitary transformation, which ensures that information is transferred to the sub-states of the environment and never to the correlations between the environment and the system.\cite{braunstein2007quantum} 

Consequently, should quantum gravity theories preserve information, an interesting case to consider is whether this information is locally conserved. In particular, concerns on whether information content is conserved in quantum gravity systems stem largely from the black hole information paradox.\cite{hawking1976breakdown, almheiri2020entropy, hawking2005information, page1993information} The most adopted solutions to the paradox employ the holographic principle (particularly, the Anti-de Sitter/Conformal Field Theory (AdS/CFT) correspondence)\cite{susskind1995world, bousso2002holographic, maldacena1999large}, which requires the quantum information of the matter and radiation degrees of freedom falling through the event horizon towards the black hole singularity be stored on the surface area. Thus, unidentified mechanisms such as quantum entanglement are thought to viably encode the quantum information on the thermal radiation of particles escaping the black hole, in a bid to guarantee unitarity of the black hole formation and evaporation process.\cite{grudka2018black, braunstein2013better, almheiri2019entropy}
However, during gravitational collapse, since the black hole horizon forms spontaneously when radius of the spherically-symmetric collapsing star becomes less than the Schwarzschild radius (it is strictly non-local\cite{booth2005black}), it would be a serendipitous result if the information they store is indeed locally conserved. 

Moreover, a heuristic argument by Susskind for constructing a Schwarzschild black hole bit by bit using quantum particles, highlights another challenge.\cite{Susskind2013} According to Susskind, it is imperative that only a bit of information per quantum particle is absorbed or emitted by the growing or shrinking black hole. For instance, considering mass-less U($1$) gauge particles, this requires that the wavelength be comparable to the Schwarzschild radius/black hole inverse temperature (speed of light, $c = 1$), $\lambda \propto 2\pi r_{\rm S} = 8\pi GM = \beta$, where $G$ is the gravitation constant, $\beta$ is the inverse temperature and $M$ is the black hole mass, thus cutting-off unnecessary modes of quantum information such as particle position from being stored by the black hole. Consequently, any emission or absorption of a single particle changes only the mass-energy of the black hole by the energy of a single particle (reduced Planck's constant, $\hbar = 1$), $\delta M = 2\pi/\lambda$, implying that the change in the Schwarzschild radius is comparable to the Compton wavelength of the black hole, $\delta r_{\rm S} = 2G\delta M = 4\pi G/\lambda \propto 4\pi G/8\pi GM = 1/2M$. 

Accordingly, the black hole surface area changes by, $\delta (4\pi r_{\rm S}^2) = 8\pi r_{\rm S}\delta r_{\rm S} \propto 16\pi GM/2M = 8\pi G$. Consequently, for a black hole built by $k$ such particles, the surface area is quantized as $4\pi r_{\rm S}^2/8\pi G \propto k \in \mathbb{Z}$.\cite{bekenstein1998quantum, mukhanov1986black, bekenstein1995spectroscopy, vaz1999mass} While this heuristic argument utilizes the de Brolie wavelength of the particle consistent with the wave-particle duality of quantum mechanics, it nonetheless poses a subtle problem to quantum field theory (the standard model of particle physics). In particular, since the particle number, comprising the black hole horizon and hence the black hole information content, may differ from conserved 
quantum numbers such as baryon number or lepton number\cite{zee2010quantum}, it may not necessarily be a conserved quantity. Nonetheless, this can be resolved by strictly considering $k$ to be related to a measure of the energy content of the black hole, which must always be conserved along geodesics since the Schwarzschild solution admits a time-like Killing vector.

Moreover, the argument not only offers crucial insights into the two relevant length scales for quantum black holes, namely the Schwarzschild radius/inverse temperature and the inverse mass/Compton wavelength, but also retains an extremely simple but desirable feature -- that the quantum state of the black hole horizon with $k$ `pixels' can be associated with the single mode quantum harmonic oscillator, $| k \rangle$ of angular frequency, $\omega = 2\pi/\lambda \propto 1/\beta$, where $[a, a^{\dagger}] = 1$, $a|k\rangle = \sqrt{k}|k - 1\rangle$ and $a^{\dagger}|k\rangle = \sqrt{k + 1}|k + 1\rangle$, with $a, a^{\dagger}$ the particle annihilation, creation operator and normalization, $\langle k|l \rangle = \delta_{kl}$. Nonetheless, since the black hole can emit and absorb the particles forming it via quantum mechanical effects such as tunneling, responsible for Hawking radiation\cite{hawking43particle, hawking1976black}, the horizon and hence the temperature is expected to likewise fluctuate. 

In this work, we formalize this heuristic argument by considering the mass-less quanta of the U($1$) gauge field to build up the black hole. We shall introduce a dimensionless proportionality constant,
\begin{align}\label{constant_eq}
    g = \lambda/\beta,
\end{align}
and define the number of pixels, $k$ on the surface area as,
\begin{align}\label{k_eq}
    k = \frac{1}{4\pi}\int d^{\,3}x \sqrt{-\det{(g_{\mu\nu})}}\mathcal{J}^0(\vec{x}), 
\end{align}
where $\nabla_{\mu}\mathcal{J}^{\mu} = 0$ and $\mathcal{J}^{\mu} = (\mathcal{J}^0, \mathcal{J}^i)$ is the locally conserved current. Thus, locally where special relativity applies, the metric tensor is Minkowski, $g_{\mu\nu} = \eta_{\mu\nu}$, while the heuristic argument
suggests we should expect that,
\begin{multline}\label{current_eq}
    \frac{\beta M}{2}\equiv \int d^{\,3}x \sqrt{-\det{(\eta_{\mu\nu})}}\mathcal{J}^{0}(\beta, \vec{x})
    \\
    = \beta\int d^{\,3}x \sqrt{-\det{(\eta_{\mu\nu})}}\langle k| (T^{00}(x) + \frac{\Lambda}{8\pi G}\eta^{00})|k \rangle\\
    = 2\times\frac{2\pi\beta g}{\lambda}\langle k |\frac{1}{2} \left( a^{\dagger}a + aa^{\dagger} \right)| k \rangle - \frac{\beta\Lambda V}{8\pi G}
    = 4\pi k,
\end{multline}
where we have accounted for the two polarization states, used eq. (\ref{constant_eq}) and $\beta = 8\pi GM$. In addition, we have set,
\begin{align}\label{V_eq}
    V = \frac{2\pi}{M\Lambda} = \int d^{\,3}x \sqrt{-\det{(\eta_{\mu\nu})}},
\end{align}
with $V$ the spatial volume, $\Lambda$ the cosmological constant, 
\begin{align}
    T^{\mu\nu} = g\left (F^{\mu\alpha}F_{\alpha}^{\,\,\nu} - \frac{1}{4}F^{\alpha\beta}F_{\alpha\beta}\eta^{\mu\nu} \right ),
\end{align}
the trace-less energy-momentum tensor of the U($1$) gauge field satisfying $\partial_{\mu}T^{\mu\nu} = 0$, $F_{\mu\nu} = \partial_{\mu}A_{\nu} - \partial_{\nu}A_{\mu}$ the field strength and $A_{\mu}$ the gauge field.

Gravitational effects are incorporated via minimal coupling by imposing the Einstein Field Equations,
\begin{align}\label{semi_classical_eq}
    R_{\mu\nu} - \frac{1}{2}Rg_{\mu\nu} + \Lambda g_{\mu\nu} =  8\pi G \langle k | T_{\mu\nu} | k \rangle,
\end{align}
on field equations of the complex-Hermitian form,\cite{kanyolo2021reproducing}
\begin{align}\label{novel_eq}
    \nabla_{\mu}\mathcal{K}^{\mu}_{\,\,\nu} = \beta\Psi^*\partial_{\nu}\Psi,
\end{align}
where $\Psi = \sqrt{\rho}\exp(iS)$ is a complex-valued function with $S$ a suitable action and $\rho$ a density function, $\mathcal{K}_{\mu\nu} = R_{\mu\nu} + i\mathcal{F}_{\mu\nu}$ is a complex-Hermitian tensor, $R_{\mu\nu}$ is the Ricci tensor and $\mathcal{F}_{\mu\nu} = M\nabla_{\mu}\xi_{\nu}$ with $\xi^{\nu}$ the time-like Killing vector. While eq. (\ref{novel_eq}) contains a complex-Hermitian tensor ($K_{\mu\nu} = R_{\mu\nu} + iF_{\mu\nu}$), its structure completely differs from complex general relativity\cite{einstein1945generalization, einstein1948generalized} since the metric tensor and affine connection are strictly real and torsion free.\cite{kanyolo2021partition} In fact, the form of eq. (\ref{novel_eq}) has been explored by the present
authors to not only consider asymptotic behavior in dark matter dominated spiral galaxies but also in other contexts of
emergent gravity in condensed matter systems such as layered materials used as battery electrodes.\cite{kanyolo2021reproducing, kanyolo2020idealised, Kanyolo2022conformal}    

The main result is that the field equations readily formalize the heuristic argument above for Schwarzchild black holes, whilst guaranteeing a locally-conserved information current given by,
\begin{subequations}
\begin{multline}\label{current_eq2}
    \mathcal{J}_{\mu} = -8\pi GM\Im{(\Psi^*\partial_{\nu}\Psi)}\\
    = -\frac{\beta}{2i}\left (\Psi^*\partial_{\mu}\Psi - \Psi (\partial_{\mu}\Psi)^* \right ) = \frac{\beta M}{2}\rho u_{\mu},
\end{multline}
exists, where $u^{\mu}$ is the four-velocity of the black hole and thus, by eq. (\ref{current_eq}), the density function, $\rho$ must be normalized as,
\begin{align}\label{normalization_eq}
    \int d^{\,3}x\,\sqrt{-\det{(g_{\mu\nu})}}\,\rho u^0 = 1,
\end{align}
\end{subequations}
where,
\begin{align}\label{Imaginary_eq}
    \nabla_{\mu}\mathcal{F}^{\mu\nu} = -\mathcal{J}^{\nu},
\end{align}
is the imaginary part of eq. (\ref{novel_eq}) and the real part corresponds to the trace of eq. (\ref{semi_classical_eq}). Crucially, the average information stored by the black hole on its surface is proportional to the entropy of the gauge particles, suggesting no information will be lost when the particles are emitted or absorbed by the black hole. Nonetheless, one needs to rigorously track the black hole formation and evaporation process for a definitive conclusion.\footnote{\textit{e.g.} as attempted in reference \citenum{almheiri2019entropy}.} Moreover, further extensive work may be needed in order to expand the formalism to apply to other black hole solutions such as the Kerr-Newman metric. 

Throughout this paper, we employ Einstein's summation convention, and natural units, where Planck's constant $\hbar$, Boltzmann constant, $k_{\rm B}$ and speed of light in vacuum $c$ are both set to unity ($\hbar = k_{\rm B} = c = 1$). We also assume a torsion-free $1 + 3$ dimensional space-time, where $\nabla_{\mu}$ is the metric compatible covariant derivative, \textit{i.e.} $\nabla_{\sigma}g_{\mu\nu} = 0$, while the metric signature convention is taken to be diag($\eta_{\mu\nu}$) $= (-1, 1, 1, 1)$ where $\eta_{\mu\nu}$ is the Minkowski metric. 

\,\,
\textit{\textbf{Einstein manifolds}} -- We shall first consider Einstein manifolds, \textit{e.g.} de Sitter vacuum, which satisfies, 
\begin{align}\label{Einstein_Manifold_eq}
    R_{\mu\nu} - \frac{1}{2}Rg_{\mu\nu} + \Lambda g_{\mu\nu} = 8\pi G\lim_{\rho \rightarrow 0}M\rho u_{\mu}u_{\nu} = 0,
\end{align}
where $\Lambda$ is the cosmological constant, $M$ is the central mass, $u_{\mu}$ is the 4-velocity vector, $\rho$ is a density function to be determined, $R_{\mu\nu} = R^{\rho}_{\,\,\mu\rho\nu}$ and $R^{\rho}_{\mu\nu\sigma}$ is the Riemann tensor satisfying,
\begin{align}\label{Riemann_eq}
    [\nabla_{\mu}, \nabla_{\nu}]u_{\sigma} = R^{\rho}_{\,\,\sigma\nu\mu}u_{\rho},
\end{align}
with $[\nabla_{\mu}, \nabla_{\nu}] = \nabla_{\mu}\nabla_{\nu} - \nabla_{\nu}\nabla_{\mu}$ the commutator and $u^{\mu}$ any four-vector (\textit{e.g.} the four-velocity) which appropriately transforms as a tensor. We shall consider eq. (\ref{Einstein_Manifold_eq}) as the constraint applied to field equations of the form,
\begin{align}
    \nabla^{\mu}\mathcal{K}_{\mu\nu} = 8\pi GM\Psi^*\partial_{\nu}\Psi,
\end{align}
also given in eq. (\ref{novel_eq}), where $\mathcal{K}_{\mu\nu} = R_{\mu\nu} + i\mathcal{F}_{\mu\nu}$ is a complex-Hermitian tensor, $g$ is a coupling constant defined in eq. (\ref{V_eq}), $i = \sqrt{-1}$, $\mathcal{F}_{\mu\nu} = M\nabla_{\mu}\xi_{\nu}$ with $\xi_{\mu}$ the time-like Killing vector, $\Psi = \sqrt{\rho}\exp(iS)$ is a complex-valued function with $S$ an action to be determined. 

Isometries on the space-time manifold guarantee that the Killing vector satisfies,\cite{thorne2000gravitation}
\begin{align}\label{Killing_eq}
    \nabla_{\mu}\xi_{\nu} = -\nabla_{\nu}\xi_{\mu}.
\end{align}
Thus, using eq. (\ref{Riemann_eq}) with $u^{\mu} \rightarrow \xi^{\mu}$ and applying eq. (\ref{Killing_eq}), we find Killing vectors must satisfy,
\begin{subequations}\label{Killing_condition_eq}
\begin{align}
    \nabla_{\mu}\nabla^{\nu}\xi_{\sigma} = R^{\rho}_{\,\,\mu\nu\sigma}\xi_{\rho},
\end{align}
and hence,
\begin{align}
    \nabla_{\mu}\nabla^{\mu}\xi_{\sigma} = R^{\rho\mu}_{\,\,\,\,\,\mu\sigma}\xi_{\rho} = -R^{\rho}_{\,\,\sigma}\xi_{\rho}.
\end{align}
Plugging in eq. (\ref{Einstein_Manifold_eq}) into eq. (\ref{Killing_condition_eq}), we arrive at,
\begin{align}\label{Killing_condition2_eq}
    \nabla_{\mu}\nabla^{\mu}\xi_{\nu} = -\Lambda\xi_{\nu}.  
\end{align}
\end{subequations}
\begin{subequations}
Essentially, eq. (\ref{Killing_condition2_eq}) requires that the imaginary part of eq. (\ref{novel_eq}) yields, 
\begin{multline}\label{Imaginary_eq2}
    8\pi GM\Im{(\Psi^*\partial_{\nu}\Psi)} = 8\pi GM\rho\partial_{\nu}S\\
    = \nabla^{\mu}\mathcal{F}_{\mu\nu} = -M\Lambda\xi_{\nu}, 
\end{multline}
while the real part, 
\begin{align}\label{Real_eq}
    \partial_{\mu}R = 8\pi GM\partial_{\mu}\rho,
\end{align}
\end{subequations}
simply corresponds to the derivative of the trace of eq. (\ref{Einstein_Manifold_eq}) (given by $R = 8\pi GM\rho + 4\Lambda$) with $u^{\mu}u_{\mu} = -1$ and $\rho \neq 0$, where we have used the Bianchi identity, 
\begin{align}\label{Bianchi_eq}
    \nabla^{\mu}R_{\mu\nu} = \frac{1}{2}\nabla_{\nu}R.
\end{align}

Furthermore, we shall consider the Einstein-Hilbert action with the cosmological constant, 
\begin{align}
    S_{\rm EH} = \frac{1}{16\pi G}\int d^{\,4}x\sqrt{-\det(g_{\mu\nu})} \left (R - 2\Lambda \right ),
\end{align}
and make a stationary phase approximation by plugging in the trace of eq. (\ref{Einstein_Manifold_eq}) to yield,
\begin{multline}
    S \equiv S_{\rm EH}\bigg\rvert_{R = 8\pi GM\rho + 4\Lambda}\\
    = \frac{M}{2}\int d^{\,4}x\sqrt{-\det(g_{\mu\nu})}\left ( \rho + \frac{\Lambda}{4\pi GM} \right ),
\end{multline}
where $S$ is defined as the action appearing in the complex-valued function, $\Psi$. Making use of the normalization condition in eq. (\ref{normalization_eq}), with $u^0 = dt/d\tau$, where $\tau$ is the proper time, we can write, 
\begin{subequations}\label{S_eq}
\begin{multline}
    S = \frac{M}{2}\int d\tau \int d^{\,3}x \sqrt{-\det(g_{\mu\nu})}\rho u^0 + S_0\\
    = M\int d\tau + S_0, 
\end{multline}
where the constant phase is given by,
\begin{multline}\label{phase_constant_eq}
    S_0 = \frac{1}{8\pi G}\int_0^{i\beta} dt\, d^{\,3}x\sqrt{-\det(g_{\mu\nu}^{\rm E})}\,\Lambda\\
    = -\frac{\beta V\Lambda}{8\pi G} = -MV\Lambda = -2\pi,
\end{multline}
with $V$ the spatial volume given in eq. (\ref{V_eq}) and $\beta = 8\pi GM$ 
chosen as the relevant cut-off time-scale based on Wick rotation, $g_{\mu\nu} \rightarrow g_{\mu\nu}^{\rm E}$ (eq. (\ref{Wick_eq})) and $g_{\mu\nu}^{\rm E}$ is the metric in Euclidean signature where we have used,
\begin{align}
    \sqrt{-\det(g_{\mu\nu})} = -i\sqrt{-\det(g_{\mu\nu}^{\rm E})}.
\end{align}
\end{subequations}

Proceeding, we can substitute eq. (\ref{S_eq}) into eq. (\ref{Imaginary_eq2}) to yield,
\begin{align}\label{lambda_rho_eq}
    \xi_{\mu} = \frac{8\pi GM}{2\Lambda}\rho u_{\mu},
\end{align}
where we have used $\int d\tau = -\int u_{\mu}dx^{\mu}$. Considering the case where,
\begin{align}\label{velocity_eq}
    u^{\mu} = \exp(\Phi)\xi^{\mu}, 
\end{align}
with $\Phi$ a scalar function, it is rather straight-forward to show that $\Phi$ is the Newtonian potential, 
\begin{subequations}
\begin{multline}\label{div_Phi_eq}
    u^{\mu}\nabla_{\mu}u_{\nu}
    = \exp(\Phi)\xi^{\mu}\nabla_{\mu}(\exp(\Phi)\xi_{\nu})\\
    = \exp(2\Phi)\xi^{\mu}\nabla_{\mu}\xi_{\nu}
    + \exp(\Phi)\xi_{\nu}\xi^{\mu}\nabla_{\mu}\exp(\Phi)\\
    = -\frac{1}{2}\exp(2\Phi)\nabla_{\nu}(\xi^{\mu}\xi_{\mu})\\
    = \frac{1}{2}\exp(2\Phi)\nabla_{\nu}\exp(-2\Phi)
    = -\nabla_{\nu}\Phi.
\end{multline}
Moreover, the conservation condition imposed by $\nabla_{\mu}\nabla_{\nu}\mathcal{F}^{\mu\nu} = 0$ requires that, $\nabla_{\mu}\left (\rho u^{\mu} \right ) = 0$, which is equivalent to the condition, 
\begin{multline}\label{rho_eq}
    \frac{d}{d\tau}\rho = u^{\mu}\nabla_{\mu}\rho = -\rho\nabla_{\mu}u^{\mu}\\
    = -\rho\exp(\Phi)\xi^{\mu}\nabla_{\mu}\Phi = -\rho u^{\mu}\nabla_{\mu}\Phi = -\rho\frac{d}{d\tau}\Phi,
\end{multline}
where we have used $\nabla_{\mu}\xi^{\mu} = 0$. This can be solved to yield,
\begin{align}
    \rho = \rho_{\rm c}\exp(-\Phi),
\end{align}
\end{subequations}
for the density function with $\rho_{\rm c}$ the proportionality constant. Moreover, eq. (\ref{V_eq}), eq. (\ref{phase_constant_eq}), eq. (\ref{lambda_rho_eq}) and eq. (\ref{velocity_eq}) require that, 
\begin{align}
    \rho_{\rm c} = \frac{\Lambda}{4\pi GM} = \frac{1}{2GM^2V}.
\end{align}
This is consistent with the normalization condition in eq. (\ref{normalization_eq}), where $\xi^{\mu} = (1, \vec{0})$ is the time-like Killing vector, when,
\begin{subequations}
\begin{align}\label{mass_eq}
    M/2\kappa = 1,
\end{align}
since we have, 
\begin{align}
    1 = \int d^{\,3}x\,\sqrt{-\det{(g_{\mu\nu})}}\,\rho u^0 = V\rho_{\rm c}. 
\end{align}
\end{subequations}
This requires that the inverse temperature, $\beta = 8\pi GM = 4\pi/M$ be proportional to the Compton wavelength, thus unifying the two relevant length scales.
Moreover, another serendipitous feature of eq. (\ref{mass_eq}) is that, for a spatial volume that extends to infinity, $V \rightarrow \infty$, the density function approaches zero, $\rho \rightarrow 0$, and we recover, from eq. (\ref{Einstein_Manifold_eq}), the Einstein vacuum condition, $R_{\mu\nu} = \Lambda g_{\mu\nu}$, as required. Moreover, since the black hole mass is $M = 2\pi/\Lambda V$ in eq. (\ref{V_eq}), keeping $M$ fixed while taking the spatial volume to infinity requires that the cosmological constant approaches zero, $\Lambda \rightarrow 0$ hence recovering $R_{\mu\nu} = 0$. Nonetheless, eq. (\ref{phase_constant_eq}) guarantees the complex-valued function, $\Psi$ is single-valued at finite $\Lambda$ and $V$. Finally, eq. (\ref{mass_eq}) is too constraining since, from the heuristic argument, we ought to expect that $\beta M/2 = 4\pi k \neq 2\pi$ violating eq. (\ref{mass_eq}), where $k \in \mathbb{Z}$ is a positive integer corresponding to the number of pixels on the black hole surface. This warrants us to incorporate vacua other than the Einstein vacuum in our formalism. 

\,\,
\textit{\textbf{Formalizing the argument}} -- Particularly, we shall consider the Einstein Field Equations, 
\begin{subequations}\label{EFE_eq}
\begin{align}
    R_{\mu\nu} - \frac{1}{2}Rg_{\mu\nu} + \Lambda g_{\mu\nu} = 8\pi G\lim_{\rho \rightarrow 0} T_{\mu\nu} \neq 0,
\end{align}
where the energy-momentum tensor is given by,
\begin{align}
    T_{\mu\nu} = M\rho u_{\mu}u_{\nu} + g\left (F^{\alpha}_{\,\,\mu}F_{\alpha\nu} - \frac{1}{4}F^{\alpha\beta}F_{\alpha\beta}g_{\mu\nu} \right ),
\end{align}
\end{subequations}
with $F_{\mu\nu} = \partial_{\mu}A_{\mu} - \partial_{\nu}A_{\mu}$ the gauge field strength and $A_{\mu}$ the U($1$) gauge potential. The bianchi identity in eq. (\ref{Bianchi_eq}) and the metricity condition, $\nabla_{\mu}g_{\nu\sigma} = 0$ requires that,
\begin{align}\label{energy_div_eq}
    \nabla_{\mu}\lim_{\rho \rightarrow 0} T^{\mu\nu} = 0,
\end{align}
which is guaranteed by the equations of motion,
\begin{align}\label{geodesic_eq}
    Mu^{\mu}\nabla_{\mu}u_{\nu} = gu^{\mu}F_{\mu\nu},
\end{align}
together with Maxwell's equations,
\begin{subequations}\label{Maxwell_eq}
\begin{align}
    \nabla_{\mu}F^{\mu\nu} = \lim_{\rho \rightarrow 0} \rho u^{\nu} = 0,
\end{align}
and the Jacobi identity,
\begin{align}
    \nabla_{\mu}F_{\nu\sigma} + \nabla_{\nu}F_{\sigma\mu} + \nabla_{\sigma}F_{\mu\nu} = 0.
\end{align}
\end{subequations}

We shall take the semi-classical approach and the limit $\rho \rightarrow 0$, to transform the Einstein Field Equations in eq. (\ref{EFE_eq}) into eq. (\ref{semi_classical_eq}) with $\lambda$ the single-mode wavelength given in eq. (\ref{constant_eq}), where a standard calculation yields, 
\begin{subequations}\label{average_T_eq}
\begin{align}
    \lim_{\rho \rightarrow 0, g_{\mu\nu} \rightarrow \eta_{\mu\nu}}\langle k | T^{00} | k \rangle = 2\times\frac{2\pi g}{\lambda}(k + 1/2)V^{-1},
\end{align}
which is satisfied locally as $g_{\mu\nu} \rightarrow \eta_{\mu\nu}$ with the factor of $2$ accounting for polarization. Here, we have used,
\begin{multline}
    T^{00} = g\sum_{i,j = 1}^{3}(\eta_{ij}\eta^{00}\eta^{00}F_{i0}F_{j0} - \frac{1}{2}\eta^{00}\eta_{ij}F_{0i}F_{0j}\eta^{00})\\
    - \frac{g}{4}\sum_{i,j, k, l = 1}^{3}(\eta_{ik}\eta_{jl}F_{ij}F_{kl}\eta^{00})
    = g\sum_{i,j = 1}^{3}\delta_{ij}\frac{1}{2}(E_iE_j + B_iB_j)\\
    = \frac{g}{V}\sum_{s = \pm}\sum_{|\vec{p^s}| = \frac{2\pi}{g\beta}}\omega_{|\vec{p^s}|}^s\left ((a_{|\vec{p^s}|}^s)^{\dagger}\,a_{|\vec{p^s}|}^s    + \frac{1}{2} \right ),
\end{multline}
\end{subequations}
where $\eta_{00} = \eta^{00} = -1$, $\eta_{0j} = \eta^{0j} = 0$, $\eta_{ij} = \eta^{ij} = \delta_{ij}$ is the Kronecker delta, $F_{0i} = E_i$ and $\frac{1}{2}\sum_{i,j = 1}^{3}\varepsilon_{ijk}F_{ij} = B_k$ have been decomposed into the sum of Fourier modes over discrete momenta, $|\vec{p^s}| = \sqrt{\sum_{i = 1}^3(p_i^s)^2} = \omega^s_{|\vec{p^s}|} = 2\pi/\lambda$ with $p_{\mu}^sp^{s\mu} = 0$, $p^{s\mu} = (\omega_{|\vec{p^s}|}^s, \vec{p^s})$ the four-momentum and $s = \pm$ the two polarization states. The Fock space of the gauge particles, $|k\rangle$ is the short hand for, 
\begin{subequations}\label{k_state_eq}
\begin{align}
    |k \rangle = \prod_{|\vec{p^s}| = \frac{2\pi}{g\beta}}\prod_{s = \pm} |k^s_{|\vec{p^s}|} \rangle = |k^+_{2\pi/\lambda}\rangle\otimes|k^-_{2\pi/\lambda}\rangle,\\
    \langle k|k \rangle = \langle k^+_{2\pi/\lambda}|k^+_{2\pi/\lambda}\rangle\otimes\langle k^-_{2\pi/\lambda}|k^-_{2\pi/\lambda}\rangle = 1,
\end{align}
\end{subequations}
where the black hole is built by an equal number of particles of each polarization state, $k = k^+_{2\pi/\lambda} = k^-_{2\pi/\lambda}$ and the creation and annihilation operators satisfy the usual quantum harmonic oscillator relations,
\begin{subequations}
\begin{align}
    a_{|\vec{p^s}|}^s|k^s_{|\vec{p^s}|} \rangle = \sqrt{k^s_{|\vec{p^s}|}}|k^s_{|\vec{p^s}|} - 1\rangle,\\
    (a_{|\vec{p^s}|}^s)^{\dagger}|k^s_{|\vec{p^s}|} \rangle = \sqrt{\left (k^s_{|\vec{p^s}|} + 1 \right )}|k^s_{|\vec{p^s}|} + 1\rangle,\\
    \left [a_{|\vec{p^r}|}^r, (a_{|\vec{q^s}|}^s)^{\dagger}\right ] = \delta_{rs}\delta_{|\vec{q^r}||\vec{p^s}|}.
\end{align}
\end{subequations}
Thus, when the momentum is considered continuous, we ought to make the identification,
\begin{subequations}
\begin{align}
   \frac{g}{V}\sum_{|\vec{p^s}| = \frac{2\pi}{g\beta}} \rightarrow 
   \frac{1}{(2\pi)^3}\int_0^{\Lambda_{\rm cut.}} 4\pi |\vec{p^s}|^2d|\vec{p^s}|,
\end{align}
in order for the thermal average to yield,
\begin{multline}
    \langle T^{00} \rangle  = \frac{g}{V}\sum_{s = \pm}\sum_{|\vec{p^s}| = \frac{2\pi}{g\beta}}\omega_{|\vec{p^s}|}^s\left (\left \langle (a_{|\vec{p^s}|}^s)^{\dagger}\,a_{|\vec{p^s}|}^s \right \rangle + \frac{1}{2} \right )\\
    =  \sum_{s = \pm}\int_0^{\Lambda_{\rm cut.}}\frac{|\vec{p^s}|^3}{2\pi^2}d|\vec{p^s}|\left (\frac{1}{\exp(\beta|\vec{p^s}|) - 1} + \frac{1}{2} \right ),\\
    = \int_0^{\Lambda_{\rm cut.}}\left (\frac{\omega}{\pi} \right )^2\left ( \frac{\omega}{\exp(\beta\omega) - 1} + \frac{\omega}{2}\right )d\omega\\
    = \frac{4\sigma(\beta\Lambda_{\rm cut.})}{\beta^4} + \frac{\Lambda}{8\pi G},
\end{multline}
\end{subequations}
where we have set $|\vec{p^{\pm}}| = \omega_{|\vec{p^{\pm}}|} = \omega$, and the cut-off energy scale and the constant $\sigma(\beta\Lambda_{\rm cut.})$ respectively to,
\begin{subequations}
\begin{align}
    \Lambda_{\rm cut.} = \left (\frac{\pi\Lambda}{G} \right )^{1/4},\\
    \sigma(\beta\Lambda_{\rm cut.}) = \int_0^{\beta\Lambda_{\rm cut.}}\frac{x^3/4\pi^2}{\exp(x) - 1}dx.
\end{align}
\end{subequations}
Consequently, since astrophysical black holes have a vanishing temperature, $\beta \rightarrow \infty$, we recover the Stephan-Boltzmann constant, $\sigma(\infty) = \pi^2/60$, as expected. 

Proceeding, since the temperature, $1/\beta$ and the wavelength $\lambda$ are both blue-shifted by the introduction of the gravitational field by equal amounts, 
\begin{subequations}\label{blue_shift_eq}
\begin{align}
    \lambda \rightarrow \lambda \sqrt{-g_{00}(x)},\\
    \beta \rightarrow \beta\sqrt{-g_{00}(x)},
\end{align}
\end{subequations}
we can adopt the condition in eq. (\ref{constant_eq}) for curved space-times without loss of generality. Moreover, to obtain the action, $S$ in eq. (\ref{novel_eq}) as before, we consider the stationary phase approximation, 
\begin{multline}\label{S_eq2}
    S \equiv S_{\rm EH}\bigg\rvert_{R = 8\pi GM\rho + 4\Lambda} + S_{\rm gauge}\bigg\rvert_{\nabla_{\mu}F^{\mu\nu} = \rho u^{\nu}}\\
    = \frac{M}{2}\int d\tau + S_0 - \frac{g}{4}\int d^{\,4}x\sqrt{-\det(g_{\mu\nu})}F_{\alpha\beta}F^{\alpha\beta}\\
    = \frac{M}{2}\int d\tau + S_0 + \frac{g}{2}\int d^{\,4}x\sqrt{-\det(g_{\mu\nu})}A_{\beta}\nabla_{\alpha}F^{\alpha\beta}\\
    = \frac{M}{2}\int d\tau + S_0 + \frac{g}{2}\int d\tau\int d^{\,3}x \sqrt{-\det(g_{\mu\nu})}(A_{\beta}u^{\beta})\rho u^0\\
    = \frac{M}{2}\int d\tau + \frac{g}{2}\int d\tau \langle A_{\alpha}u^{\beta} \rangle + S_0 ,
\end{multline}
where we have discarded the boundary term, 
\begin{subequations}
\begin{align}
    \int d^{\,4}x\sqrt{-\det(g_{\mu\nu})}\nabla_{\alpha}(A_{\beta}F^{\alpha\beta}) = 0, 
\end{align}
and defined the average operation as,
\begin{align}\label{average_eq}
    \langle \cdots \rangle = \int d^{\,3}x \sqrt{-\det(g_{\mu\nu})}(\cdots)\rho u^0. 
\end{align}
\end{subequations}

Thus, we shall set,
\begin{align}\label{SV_eq}
   \left \langle \frac{g}{2}\int d\tau \langle k|A_{\alpha}u^{\alpha}|k \rangle \right \rangle + S_0 = 2\pi \nu(k),
\end{align}
in eq. (\ref{S_eq2}) with $\nu(k) \in \mathbb{Z}$ is some integer that depends on $k$, in order to be consistent with eq. (\ref{current_eq2}). Evidently,
\begin{subequations}
\begin{multline}
     \int d^{\,3}x \sqrt{-\det{(g_{\mu\nu})}}\mathcal{J}^0(\vec{x})\\
     = \frac{\beta M}{2}\int d^{\,3}x \sqrt{-\det{(g_{\mu\nu})}}\rho u^0 = \frac{\beta M}{2}, 
\end{multline}
where eq. (\ref{normalization_eq}) has been applied. Since eq. (\ref{novel_eq}) is invariant under the introduction of a trace-less energy-momentum tensor in Einstein Field Equations, this implies that $\Psi$ ought to be single-valued, in accordance with eq. (\ref{SV_eq}). Furthermore, by the imaginary part of eq. (\ref{novel_eq}) given in eq. (\ref{Imaginary_eq}), 
\begin{multline}\label{betaM_eq}
   \frac{\beta M}{2} =  \int d^{\,3}x \sqrt{-\det{(g_{\mu\nu})}}\mathcal{J}^0(\vec{x})\\
    = -\int d^{\,3}x \sqrt{-\det{(g_{\mu\nu})}}\nabla_{\mu}\mathcal{F}^{\mu0}\\
    = -M\int d^{\,3}x \sqrt{-\det{(g_{\mu\nu})}}\nabla_{\mu}\nabla^{\mu}\xi^0\\
    = M\int d^{\,3}x \sqrt{-\det{(g_{\mu\nu})}}R^{\mu 0}\xi_{\mu}\\
    = 8\pi GM\int d^{\,3}x \sqrt{-\det{(g_{\mu\nu})}}\langle k|T^{00}|k \rangle\\
    + \frac{2\pi}{V}\int d^{\,3}x \sqrt{-\det{(g_{\mu\nu})}}g^{00},
\end{multline}
\end{subequations}
where we have used $V = 2\pi/M\Lambda$ from eq. (\ref{V_eq}). Consequently, we recover eq. (\ref{current_eq}) in the limit $g_{\mu\nu} \rightarrow \eta_{\mu\nu}$ where we have used eq. (\ref{average_T_eq}). For astrophysical black holes, $k \gg 1/2$, obtaining $k + 1/2 \simeq k$, where the particle number dominates the zero point term. Since $\Lambda$ is defined such that it cancels out the zero point term, we can reasonably set, $\Lambda \rightarrow 0$, without loss of consistency. 

Finally, plugging in eq. (\ref{velocity_eq}), where $\xi^{\mu} = (1, \vec{0})$ is the time-like Killing vector, into the time component of the imaginary part of eq. (\ref{novel_eq}) with $g_{\mu\nu} = \eta_{\mu\nu}$ requires the expression for the proportionality constant, $\rho_{\rm c}$ in eq. (\ref{rho_eq}) to satisfy,
\begin{align}\label{recover_eq}
    \frac{4\pi k}{V} = \frac{8\pi GM^2}{2}\rho u^0 = \frac{\beta M}{2}\rho_{\rm c}.
\end{align}
Moreover, plugging in eq. (\ref{velocity_eq}) and eq. (\ref{rho_eq}) into eq. (\ref{normalization_eq}) yields and expression for the proportionality constant, $\rho_{\rm c} = 1/V$ thus recovering eq. (\ref{current_eq}) from eq. (\ref{recover_eq}). For infinite volume, $V \rightarrow \infty$ at fixed $M$, $\rho_{\rm c} \rightarrow 0$, hence guaranteeing the limits, $\rho \rightarrow 0$ and $\Lambda \rightarrow 0$ (eq. (\ref{V_eq})). Lastly, by eq. (\ref{div_Phi_eq}) and eq. (\ref{geodesic_eq}), we can write,
\begin{subequations}
\begin{align}
    \frac{g}{M}u^{\mu}F_{\mu\nu} = -\partial_{\nu}\Phi,
\end{align}
which can be rearranged using eq. (\ref{V_eq}) and eq. (\ref{velocity_eq}) to yield,
\begin{align}\label{nabla_rho_eq}
    \partial_{\nu}\rho = -\rho\partial_{\nu}u^0 = \frac{g\Lambda}{2\pi}F_{0\nu}. 
\end{align}
\end{subequations}

\,\,
\textit{\textbf{Quantum effects}} -- We begin by noting the fact that the Lie derivative of the Ricci scalar, $R$ (and equivalently, via eq. (\ref{novel_eq}), the density function, $\rho$) in the direction of the Killing vector identically vanishes,
\begin{multline}\label{no_real_eq}
    0 = \nabla_{\mu}\nabla_{\nu}\nabla^{\mu}\xi^{\nu} = \nabla_{\mu}(R^{\mu\nu}\xi_{\nu})\\
    = \frac{1}{2}\nabla^{\nu}R = 4\pi GM\xi_{\nu}\nabla^{\nu}\rho,
\end{multline}
and hence, due to the anti-symmetry, $F_{\mu\nu} = -F_{\nu\mu}$, it is consistent with eq. (\ref{nabla_rho_eq}). Moreover, observe that, by eq. (\ref{normalization_eq}) and eq. (\ref{average_eq}), $\langle 1 \rangle = 1$ whereas,
\begin{multline}\label{p_eq}
   p^{\mu} = M\int d^{\,3}x \sqrt{-\det(g_{\mu\nu})}\rho u^0u^{\mu}\\
   = M\langle u^{\mu} \rangle = \int d^{\,3}x \sqrt{-\det(g_{\mu\nu})}t^{0\mu},
\end{multline}
is the momentum calculated from the energy-momentum tensor, $t^{\mu\nu} = M\rho u^{\mu}u^{\nu}$. Consequently, to make it apparent that $\Psi$ is a quantum mechanical wave function, we shall introduce the definitions,
\begin{subequations}\label{Psi_def_eq}
\begin{align}
    \Psi(\vec{x}, t) = \exp\left (-i\frac{M}{2}u_0(\vec{x})t \right )\Psi(\vec{x}, 0),\\
    \Psi(\vec{x}, 0) = \sqrt{\rho(\vec{x})}\exp\left (-i\frac{M}{2}\int\vec{u}(\vec{x})\cdot d\vec{x} \right ),
\end{align}    
and,
\begin{align}
    \Psi(\vec{x}, 0) = \langle \vec{x}|\Psi \rangle,\\
    \Psi^*(\vec{x}, 0) = \langle \Psi|\vec{x} \rangle,
\end{align}
\end{subequations}
where, $u_{\mu} = g_{\mu\nu}u^{\nu} = (u_0(\vec{x}), \vec{u}(\vec{x}))$ is the covariant four-velocity related to the contravariant four-velocity given in eq. (\ref{velocity_eq}) and the space-time manifold strictly admits a time-like Killing vector satisfying $\xi^{\mu}\partial_{\mu}g_{\alpha\beta} = \partial g_{\mu\nu}/\partial t = 0$, and hence by eq. (\ref{nabla_rho_eq}), $\partial u_0/\partial t = g_{0\mu}\partial u^{\mu}/\partial t = g_{00}\partial u^0/\partial t = 0$, in order to write $\Psi(\vec{x}, t) = \sqrt{\rho(\vec{x})}\exp(iS(\vec{x}, t))$ with $S = (M/2)\int d\tau = -(M/2)\int u_{\mu}dx^{\mu}$ from eq. (\ref{Psi_def_eq}). Moreover, 
\begin{align}
    \hat{1} = \int d^{\,3}x\sqrt{-\det{(g_{\mu\nu})}}u^0(\vec{x})|\vec{x}\rangle \langle \vec{x}|,
\end{align}
which guarantees the wave function, $|\Psi \rangle$ is appropriately normalized,
\begin{multline}
    \langle \Psi|\Psi \rangle = \langle \Psi|\hat{1}|\Psi \rangle\\
    = \int d^{\,3}x\sqrt{-\det{(g_{\mu\nu})}}u^0(\vec{x})\langle \Psi|\vec{x}\rangle \langle \vec{x}|\Psi \rangle\\
    = \int dV u^{0}|\Psi|^2 = 1,
\end{multline}
as expected, where the volume element, $V$ is given in eq. (\ref{V_eq}). Consequently, we find, 
\begin{multline}
    2i\xi^{\mu}\langle \Psi|\partial_{\mu}|\Psi \rangle
    = \int d^{\,3}x \sqrt{-\det(g_{\mu\nu})}u^0i\xi^{\mu}\Psi^*\partial_{\mu}\Psi\\
    = \int d^{\,3}x \sqrt{-\det(g_{\mu\nu})}\xi^{\mu}\rho u^0u_{\mu}\\
    = \int d^{\,3}x \sqrt{-\det(g_{\mu\nu})}\xi_{\mu}t^{\mu 0} = \xi_{\mu}p^{\mu},
\end{multline}
where $\xi^{\mu}$ is any Killing vector, and we have used eq. (\ref{novel_eq}), eq. (\ref{p_eq}) and eq. (\ref{no_real_eq}). Proceeding, we consider the operators, 
\begin{subequations}
\begin{align}
    p_{\bar{a}} = 2i\xi^{\mu}_{\,\,\bar{a}}\partial_{\mu} \equiv i\partial_{\bar{a}},\\
    x_{\bar{a}} = \xi_{\mu\bar{a}}x^{\mu},
\end{align}
\end{subequations}
where $\bar{a} = 0, 1, 2, 3$ labels the space/time like Killing vector and $\xi_{\mu\bar{a}} = g_{\mu\nu}\xi^{\nu}_{\,\,\bar{a}}$, whose commutation relations yield, 
\begin{align}
    [p_{\bar{a}},x_{\bar{b}}] = 2i\xi^{\mu}_{\,\,\bar{a}}\xi_{\mu\bar{b}}, 
\end{align}
implying a modified uncertainty relation by the 
gravitational field. For instance, for a space time with two Killing vectors oriented along the $0, 3$ directions, respectively corresponding to a time-like, $\xi^{\mu}_{\,\,\bar{0}} = (1, \vec{0})$, and space-like Killing vector, $\xi^{\mu}_{\,\,\bar{3}} = (\vec{0}, 1)$, the commutation relations reduce to,
\begin{subequations}
\begin{align}
    [p_{\bar{0}},x_{\bar{0}}] = 2i\xi^{\mu}_{\,\,\bar{0}}\xi_{\mu\bar{0}} = 2ig_{00},\\
    [p_{\bar{3}},x_{\bar{3}}] = 2i\xi^{\mu}_{\,\,\bar{3}}\xi_{\mu\bar{3}} = 2ig_{33},
\end{align}
and,
\begin{multline}
    [p_{\bar{0}},x_{\bar{3}}] = 2i\xi^{\mu}_{\,\,\bar{0}}\xi_{\mu\bar{3}} = 2ig_{03}\\
    = 2ig_{30} = 2i\xi^{\mu}_{\,\,\bar{3}}\xi_{\mu\bar{0}} = [p_{\bar{3}},x_{\bar{0}}].
\end{multline}
\end{subequations}

\textit{\textbf{Geometric phase}} -- We wish to introduce periodicity to the time coordinate in order to define a geometric phase analogous to the Berry phase\cite{berry1984quantal, cohen2019geometric}, with properties where the number of particles is the first Chern number.\cite{chern1946characteristic, milnor2016characteristic} This will be achieved by working with a periodic imaginary time acquired by a Wick rotation/analytic continuation, $t = it_{\rm E}$. This changes the space-time metric with a Lorenzian signature, $g_{\mu\nu}$ to a metric with Euclidean signature, $g_{\mu\nu} \rightarrow g_{\mu\nu}^{\rm E}$. To account for the crucial change of sign in the formalism such as in eq. (\ref{betaM_eq}), we shall also require the cosmological constant to transform as $\Lambda \rightarrow -\Lambda$ (from de Sitter to anti-de Sitter) under Wick rotation, where $\Lambda \geq 0$. While this transformation is typically not carried out in standard Euclidean path integral, it ensures that the vacuum energy contributions of the gauge field can always be appropriately canceled out by the contribution of the cosmological constant.

Moreover, periodicity of Euclidean time implies that the time coordinate is periodic in $i\beta$,
\begin{align}\label{Wick_eq}
    t \rightarrow t + i\beta,
\end{align}
and the wave function in eq. (\ref{Psi_def_eq}) transforms accordingly as, 
\begin{subequations}\label{Eucl_psi_eq}
\begin{align}
    \Psi(\vec{x}, t) \rightarrow \exp\left (\frac{\beta M}{2}u_0 \right )\Psi(\vec{x}, t),\\
    \Psi^*(\vec{x}, t) \rightarrow \exp\left (-\frac{\beta M}{2}u_0 \right )\Psi^*(\vec{x}, t).
\end{align}
\end{subequations}
Thus, for quantum field theory on the space-time with operators, $\mathcal{O}_j(\vec{x}, t_{\rm E}) = \exp(\mathcal{H}t_{\rm E})\mathcal{O}_j(\vec{x}, 0)\exp(-\mathcal{H}t_{\rm E})$ where $\mathcal{H}$ is the Hamiltonian of the theory, the thermal average is given by,
\begin{multline}
    \overline{\mathcal{O}_j(\vec{x}, t_{\rm E})} =\\
    \mathcal{Z}^{-1}{\rm Tr}\left ( \exp(-\beta \mathcal{H})\mathcal{U}^{-1}(t_{\rm E})\mathcal{O}_j(\vec{x}, 0)\mathcal{U}(t_{\rm E})\right ),
\end{multline}
which is invariant under eq. (\ref{Wick_eq}), where the Euclidean time is defined as, $t_{\rm E} = it$, $\mathcal{Z} = {\rm Tr}(\exp(-\beta\mathcal{H}))$ is the partition function and $\mathcal{U}^{-1}(t_{\rm E}) = \exp(\mathcal{H}t_{\rm E}), \mathcal{U}(t_{\rm E}) = \exp(-\mathcal{H}t_{\rm E})$. Likewise, by virtue of eq. (\ref{Eucl_psi_eq}), the additional averaging by $\Psi$ also satisfies this invariance,
\begin{align}
    \langle \overline{\mathcal{O}_j(\vec{x}, t_{\rm E})} \rangle =
    \int dV u^0(\vec{x})\Psi^*(\vec{x}, t)\overline{\mathcal{O}_j(\vec{x}, t_{\rm E})}\Psi(\vec{x}, t),
\end{align}
provided, $[u_0(\vec{x}), \mathcal{O}_j(\vec{x}, t_{\rm E})] = 0$. 

Proceeding, we shall make the ansatz that, the spatial coordinates evolve with time at finite temperature as follows, 
\begin{multline}
    \vec{x} \rightarrow \overline{\vec{x}(t_{\rm E} = it)} = \\
    \mathcal{Z}^{-1}{\rm Tr}\left ( \exp(-\beta \mathcal{H})\exp(\mathcal{H}t_{\rm E})\vec{x}(0)\exp(-\mathcal{H}t_{\rm E})\right ), 
\end{multline}
where the wave function transforms as, 
\begin{align}
    |\Psi(\vec{x}, t)\rangle \rightarrow |\Psi(\vec{x}(t), 0) \rangle,
\end{align}
Now, introducing a closed path, $\mathcal{C}$, with a cut-off time scale equal to a time period given by $i\beta = i8\pi GM$, we can calculate the geometric phase, 
\begin{subequations}\label{geometric_phase_eq}
\begin{multline}
    \int d\tau\langle \Psi(\vec{x}, t)|\partial_{\bar{0}}|\Psi(\vec{x}, t) \rangle\\
    = \int d\tau\int d^{\,3}x\sqrt{-\det(g_{\mu\nu})}u^0\Psi^*\xi^{\mu}_{\,\,\bar{0}}\partial_{\mu}\Psi\\
    = \frac{1}{\beta}\int d\tau\int d^{\,3}x\sqrt{-\det(g_{\mu\nu})}u^0\xi^{\mu}_{\,\,\bar{0}}\nabla^{\nu}\mathcal{K}_{\nu\mu}\\
   = i\frac{1}{\beta}\int d\tau\int d^{\,3}x\sqrt{-\det(g_{\mu\nu})}u^0\xi^{\mu}_{\,\,\bar{0}}\nabla^{\nu}\mathcal{F}_{\nu\mu}\\
   = -i\frac{1}{\beta}\int d\tau\int d^{\,3}x\sqrt{-\det(g_{\mu\nu})}u^0\xi^{\mu}_{\,\,\bar{0}}\mathcal{J}_{\mu},\\
   = -i\frac{M}{2}\int_0^{i\beta} dt\int d^{\,3}x\sqrt{-\det(g_{\mu\nu})}\rho g_{0\mu}u^{\mu},
\end{multline}
where we have used $u^0 = dt/d\tau$ and employed eq. (\ref{no_real_eq}), which guarantees the real part of eq. (\ref{novel_eq}) vanishes. Thus, as $g_{\mu\nu} \rightarrow \eta_{\mu\nu}$, $g_{\mu 0} \rightarrow (-1, \vec{0})$ and we obtain, 
\begin{multline}
    \frac{\beta M}{2} = -\int d\tau \langle \Psi(\vec{x}(t_{\rm E}), 0)|\partial_{\bar{0}}|\Psi(\vec{x}(t_{\rm E}), 0) \rangle\\
    = \int dt_{\rm E}\int dV\Psi^*(\vec{x}(t_{\rm E}), 0)i\partial_{t_{\rm E}}\Psi(\vec{x}(t_{\rm E}), 0)\\
    = \int dV\int_{\mathcal{C}} d\vec{x}(t_{\rm E})\cdot\Psi^*(\vec{x}(t_{\rm E}), 0)i\vec{\nabla}_{\vec{x}(t_{\rm E})}\Psi(\vec{x}(t_{\rm E}), 0)\\
    = \int_{\mathcal{C}} \vec{a}(\vec{x}(t_{\rm E}))\cdot d\vec{x}(t_{\rm E}) = \int_{\mathcal{S}} \vec{b}(\vec{x}(t_{\rm E}))\cdot\vec{n}\, d^{\,2}x(t_{\rm E})\\
    = 4\pi k,
\end{multline}
\end{subequations}
where, 
\begin{subequations}\label{Berry_eq}
\begin{align}
    \vec{a}(\vec{x}(t_{\rm E})) = i\int dV\Psi^*\vec{\nabla}_{\vec{x}(t_{\rm E})}\Psi\\
    \vec{b}(\vec{x}(t_{\rm E})) = \vec{\nabla}_{\vec{x}(t_{\rm E})}\times\vec{a}(\vec{x}(t_{\rm E})),
\end{align}
\end{subequations}
are the Berry connection and curvature respectively, $\vec{\nabla}_{\vec{x}(t_{\rm E})} = \partial/\partial x^j(t_{\rm E})$, $\vec{n}$ is a vector normal to the enclosed 2 dimensional surface, $S$ by the curve, $\mathcal{C} = \partial \mathcal{S}$, and $\nu = 2k \in \mathbb{Z}$ is the first Chern number. 

This can be reconciled with eq. (\ref{SV_eq}) by recognizing that, $g_{\mu\nu} \rightarrow g_{\mu\nu}^{\rm E} \rightarrow \eta_{\mu\nu}^{\rm E}$ in Euclidean time and,
\begin{multline}
    \langle k| S_{\rm gauge} |k \rangle + S_0\\
    = \left \langle \frac{g}{2}\int d\tau\langle k| A_{\mu}(\tau)u^{\nu}(\tau) |k \rangle \right \rangle + S_0\\
    = \frac{g}{2}\int d\tau \langle k |A_{\mu}(\tau)|k \rangle u^{\nu}(\tau)\langle 1 \rangle + S_0\\
    = \frac{g}{2}\int \langle k |A_{\mu}(\tau)|k \rangle dx^{\mu}(\tau) + S_0\\
    =  -\frac{g}{4}\int_0^{i\beta} dt\int d^{\,3}x\sqrt{-\det(\eta_{\mu\nu}^{\rm E})}\langle k|F_{\rm E}^{\mu\nu}F_{\mu\nu}|k\rangle + S_0\\
    = -ig\beta\int d^{\,3}x\sqrt{-\det(\eta_{\mu\nu}^{\rm E})}\sum_{i,j = 1}^{3}\delta_{ij}\langle k |\frac{1}{2}(E_iE_j + B_iB_j)| k\rangle\\
    + S_0 = \beta V\langle k|T^{00} - \frac{\Lambda}{8\pi G}\eta_{\rm E}^{00}|k \rangle = 4\pi k = 2\pi \nu(k),
\end{multline}
where $F_{\rm E}^{\mu\nu} = \eta_{\rm E}^{\mu\alpha}\eta_{\rm E}^{\nu\beta}F_{\alpha\beta}$ and we have used $\eta_{\rm E}^{00} = \eta^{\rm E}_{00} = 1$, $\eta_{0j}^{\rm E} = \eta_{\rm E}^{0j} = 0$ and 
$\eta_{ij}^{\rm E} = \eta_{\rm E}^{ij} = \delta_{ij}$. This implies that, in eq. (\ref{Berry_eq}), $a_{\mu}(\tau) = g\langle k|A_{\mu}(\tau)| k \rangle$, where $A_{\mu}(\tau) = (0, \vec{A}(\tau))$ is the choice of gauge and $\nu(k) = 2k$ is the first Chern number.

Generally, we can take the integration surface to be the two-sphere, $\partial B = \mathcal{S} = S^2$, assumed to be the surface of the black hole, $B$ at the event horizon with coordinates $\vec{x}(t_{\rm E})$, and introduce the $p$ forms, $b = da$ where $a = a_{\mu}(x)dx^{\mu}$ is the connection $p = 1$ form. Thus, the particle number $\nu = 2k \in \mathbb{Z}$ becomes the winding number,
\begin{multline}
    \int_{\mathcal{S} = S^2} \vec{b}(\vec{x}(t_{\rm E}))\cdot\vec{n}\, d^{\,2}x(t_{\rm E})\\
    = \int_{S^2}da = \int_{\partial S^2_+}a_+ + \int_{\partial S^2_-}a_-\\
    = \int_{S^1}(a_+ - a_-) = 2\pi \int_0^{\nu} d\theta = 2\pi \nu = 4\pi k, 
\end{multline}
where $S^2 = S^2_+ + S^2_-$ with $S^2_{+/-}$ the northern/southern hemispheres bounded by a one-sphere $\partial S^2_+ = -\partial S^2_- = S^1$ with opposite orientation while the two Berry connections $a_+, a_-$ are related by gauge invariance, $a_+ = a_- + 2\pi d\theta$. Since $\partial S^2 = \partial S^2_+ + \partial S^2_- = S^1 - S^1 = 0$, the non-trivial topology of $S^2$ guarantees the geometric phase is finite and proportional to the winding number. Thus, it is clear that, using the Schwarzschild radius, $r_{\rm S} = 2GM$, the Schwarzschild black hole information corresponds to,
\begin{align}\label{entropy_eq}
    \mathcal{I} = \frac{4\pi r_{\rm S}^2}{4G} = \frac{\beta M}{2} = 4\pi k,
\end{align}
which is the geometric phase. Finally, we can conclude that, by virtue of eq. (\ref{k_eq}) and eq. (\ref{Imaginary_eq}), black hole information is locally conserved due to space-time isometries (time-like Killing vector, $\xi^{\mu} = (1, \vec{0})$) on the space-time manifold which couple to the information current, $\mathcal{J}^{\mu}$. 

\textit{\textbf{Discussion}} -- We have consistently introduced a geometric phase corresponding to the information content in Schwarzschild black holes. This quantity appropriately satisfies a local conservation law subject to minimal coupling, with other desirable properties such as the quantization of the black hole horizon in units of Planck area. The local conservation law is imposed by field equations, which not only contain the trace of Einstein Field Equations, but also a complex-valued function with properties analogous to the quantum-mechanical wave function. 

The geometric phase is defined only with solutions of Einstein Field Equations which admit a time-like Killing vector. However, since the black hole formation and evaporation process renders the space-time dynamical (implying a time dependent black hole mass and hence no time-like Killing vector exists), this poses a challenge extending the formalism herein to capture the entire black hole formation and evaporation process to conclusively resolve the black hole information paradox. This apparent issue with our formalism can also be understood quantum mechanically since the geometric phase is an adiabatic invariant describing the evolution of a quantum system restricted in its energy eigenstate despite a periodic time-evolution. In particular, for the black hole, the periodic time evolution arises from the analytic continuation of real to imaginary time, whereas the adiabatic condition is enforced by the time-like Killing vector. 

Consequently, to define black hole information as the geometric phase of some evolving quantum mechanical system at some finite temperature, the adiabatic condition requires the energy at each instance of the formation and evaporation process in real time to be approximated by a time-independent black hole mass interpreted as the relevant energy eigenstate. Further discussion on the dynamical case has been included in the appendix.

\textit{\textbf{Conclusion}} -- We expect our approach sheds light not only on the nature of black hole information, but also the limited success herein attests to the utility of the proposed field equations in capturing information theoretic aspects of quantum gravity.

\textit{\textbf{Acknowledgments}} -- The authors would like to acknowledge the financial support of TEPCO Memorial Foundation, Japan Society for the Promotion of Science (JSPS KAKENHI Grant Numbers 19K15685 and 21K14730) and Japan Prize Foundation. The authors also acknowledge fruitful discussions with D. Ntara during the cradle of the ideas herein, and especially the rigorous proofreading work on the manuscript done by Edfluent. Both authors are grateful for the unwavering support from their family members (T. M.: Ishii Family, Sakaguchi Family and Masese Family; G. M. K.: Ngumbi Family). 

\bibliography{information}

\appendix*

\section{Appendix}

\textit{\textbf{Towards Kerr-Newman black holes}} -- To extend the formalism to the Kerr-Newman black holes, a suitable place to start is the expression for the black hole mass\cite{bardeen1973four},
\begin{subequations}\label{M_Kerr_eq}
\begin{align}
    M = \frac{\kappa A}{4\pi G} + 2\Omega_{\rm B}L_{\rm B} + 2q\phi,
\end{align}
where the surface gravity, $\kappa$, the electric potential, $\phi$ and angular frequency, $\Omega_{\rm B}$ are given by, 
\begin{align}\label{Kerr_Newmann_solutions}
    \kappa = \frac{4\pi}{A}\left ( r_+ - r_{\rm S}/2 \right ),\,\,
    \phi = \frac{q}{A\varepsilon_0}r_+,\,\,
    \Omega_{\rm B} = \frac{4\pi}{A}a,
\end{align}
\end{subequations}
with $r_+ = r_{\rm S}/2 + \sqrt{(r_{\rm S}/2)^2 - a^2 - Gq^2/4\pi \varepsilon_0}$ the radius of the outer horizon, $r_{\rm S} = 2GM$ the Schwarzschild radius, $A = 4\pi r_+r_{\rm S}$ the black hole surface area, $r_{\rm S} = 2GM$ the Schwarzschild radius, $q$ the electric charge, $\varepsilon_0$ permittivity of free space, $a = L_{\rm B}/M$,
\begin{align}\label{L_eq}
    L_{\rm B} = -\frac{1}{8\pi G}\int_{\partial B} d\Sigma_{\mu\nu} \nabla^{\mu}\xi_{\bar{4}}^{\nu}, 
\end{align}
the angular momentum of the black hole and $d\Sigma_{\mu\nu} = \frac{1}{2}(\xi_{\mu}n_{\nu} - \xi_{\nu}n_{\mu})dA$ is the black hole surface element. Here, the Killing vector orthogonal to the horizon, $\xi^{\mu}\xi_{\mu} = 0$ is given (instead of $\xi^{\mu} = \xi^{\mu}_{\bar{0}}$) by, 
\begin{align}\label{sum_xi_eq}
    \xi^{\mu} = \xi^{\mu}_{\bar{0}} + \Omega_{\rm B}\xi^{\mu}_{\bar{4}} = \kappa^{-1}\xi^{\nu}\nabla_{\nu}\xi^{\mu},
\end{align}
and $\nu^{\mu}$ is the other null vector orthogonal to $\partial B$ satisfying $\xi^{\mu}n_{\mu} = 1$. 

To accurately utilize our formalism, it is prudent to multiply eq. (\ref{M_Kerr_eq}) by $4\pi GM$ and rearrange it as, 
\begin{multline}\label{non_vanish_eq}
    4\pi k - M\kappa A = 4\pi GM^2 - M\kappa A\\
     =  8\pi GM\Omega_{\rm B}L_{\rm B} + 8\pi GMq\phi \neq 0,
\end{multline}
where the non-vanishing terms arise not only from the finite electromagnetic energy of the charged black hole, but also due to the fact that the black hole is rotating. 

Note that, by a similar calculation to eq. (\ref{S_eq2}), the action in $\Psi$ ought to transform as,
\begin{subequations}
\begin{align}
    S \rightarrow S - \frac{q}{2}\int d\tau \tilde{A}_{\mu}u^{\mu}, 
\end{align}
when, 
\begin{align}
    \nabla_{\mu}F^{\mu\nu} = \frac{q}{\varepsilon_0}\rho u^{\nu},\\
    T_{\mu\nu} \rightarrow T_{\mu\nu} - \varepsilon_0\left (\tilde{F}^{\alpha}_{\,\,\mu}\tilde{F}_{\alpha\nu} - \frac{1}{4}\tilde{F}^{\alpha\beta}\tilde{F}_{\alpha\beta}g_{\mu\nu} \right ),
\end{align}
\end{subequations}
where $\tilde{F}_{\mu\nu} = \partial_{\mu} \tilde{A}_{\nu} - \partial_{\nu}\tilde{A}_{\mu}$ is the electromagnetic tensor and we have assumed, $\langle \tilde{A}_{\mu}(\tau)u^{\mu}(\tau) \rangle = \tilde{A}_{\mu}(\tau)u^{\mu}(\tau)$.

Nonetheless, we shall still require $\mathcal{F}_{\mu\nu} = M\nabla_{\mu}\xi_{\nu\bar{0}}$ in eq. (\ref{novel_eq}). Thus, introducing the volume element, $d\Sigma_{\mu} = n_{\mu}dV$, where $n_{\mu} = (1, \vec{0})$, we can write,
\begin{subequations}\label{stokes'_eq}
\begin{multline}\label{stokes'_eq1}
    4\pi GM\left ( M - q\int_B dV\Psi^*\tilde{A}^0\Psi \right )\\
    = 8\pi GMi\int_B d\Sigma_{\mu}g^{\mu\nu}\Psi^*\partial_{\nu}\Psi\\
    = i\frac{1}{2}\int_B dVg^{0\nu}\partial_{\nu}R - M\int_B d\Sigma_{\mu}g^{\mu\nu}\nabla_{\alpha}\nabla^{\alpha}\xi_{\bar{0}\nu}.  
\end{multline}
The imaginary part can be evaluated in the Boyer-Lindquist coordinates\cite{boyer1967maximal}, where $g^{0\nu}\partial_{\nu} = g^{00}\xi_{\bar{0}}^{\nu}\partial_{\nu} + g^{03}\xi_{\bar{3}}^{\nu}\partial_{\nu}$, which guarantees it identically vanishes since, $\xi^{\mu}_{\bar{a} = 0, 3}\partial_{\mu}R = 0$ by eq. (\ref{no_real_eq}). Moreover, applying Stoke's theorem to the last term yields, 
\begin{multline}\label{Stokes'_eq2}
    - M\int_B d\Sigma_{\mu}\nabla_{\nu}\nabla^{\nu}\xi_{\bar{0}}^{\mu} = -M\int_{\partial B} d\Sigma_{\mu\nu}\nabla^{\nu}\xi_{\bar{0}}^{\mu}\\
    = M\int_{\partial B} d\Sigma_{\mu\nu}\nabla^{\mu}\xi^{\nu} - M\Omega_{\rm B}\int_{\partial B} d\Sigma_{\mu\nu}\nabla^{\mu}\xi_{\bar{3}}^{\nu}\\
    = M\int_{\partial B} \kappa dA + 8\pi GM\Omega_{\rm B}L_{\rm B},
\end{multline}
\end{subequations}
where we have used eq. (\ref{L_eq}), eq. (\ref{sum_xi_eq}) and $\xi^{\mu}n_{\mu} = 1$. Lastly, plugging in eq. (\ref{Stokes'_eq2}) into eq. (\ref{Stokes'_eq2}) and using the fact that the surface gravity, $\kappa$ is constant over the horizon, $\partial B$, we reproduce eq. (\ref{non_vanish_eq}) when,
\begin{align}
    \phi = \frac{1}{2}\int_B dV\Psi^*\tilde{A}^0\Psi. 
\end{align}
However, this does not reproduce the quantized surface area as required.

Alternatively, we ought to define an effective mass and inverse temperature respectively,
\begin{subequations}
\begin{align}
    M_{\rm eff} = \frac{\sqrt{r_+r_{\rm S}}}{2G},\\
    \beta = \frac{2\pi}{\kappa} = \frac{\lambda}{g},\\
    \beta_{\rm eff} = 8\pi GM_{\rm eff},
\end{align}
\end{subequations}
where the novel field equations are given by,
\begin{subequations}\label{novel_eff_eq}
\begin{align}
    \nabla^{\mu}\mathcal{K}_{\mu\nu} = \beta_{\rm eff} \Psi_{\rm eff}^*\tilde{D}_{\nu}\Psi_{\rm eff},\\
    \mathcal{K}_{\mu\nu} = R_{\mu\nu} + i\mathcal{F}_{\mu\nu},\\
    \mathcal{F}_{\mu\nu} = \frac{\beta}{8\pi G}\nabla_{\mu}\xi_{\nu},
\end{align}
\end{subequations}
where, $\Psi_{\rm eff} = \sqrt{\rho}\exp(iS_{\rm eff})$ is the wave function, $S_{\rm eff} = (M_{\rm eff}/2)\int d\tau$ is the effective action, $\tilde{D}_{\nu} = \partial_{\mu} + q\tilde{A}_{\mu}/2$ is a gauge-covariant derivative, $\xi^{\mu}$ is strictly given by eq. (\ref{sum_xi_eq}). While appearing ad hoc, the re-definitions in eq. (\ref{novel_eff_eq}) reproduce eq. (\ref{novel_eq}) and hence the desired results in the Schwarzschild limit, $\Omega_{\rm B} \rightarrow 0, L_{\rm B} \rightarrow 0$ and $q \rightarrow 0$, since $\xi^{\mu} \rightarrow \xi^{\mu}_{\bar{0}}$, $r_+ \rightarrow r_{\rm S} = 2GM$, $M_{\rm eff} \rightarrow M$ and $\kappa \rightarrow 1/2r_{\rm S}$. Thus, re-performing the calculations in eq. (\ref{stokes'_eq}) yields, 
\begin{multline}
    \mathcal{I} = \frac{\beta_{\rm eff} M_{\rm eff}}{2}\\
    = i\beta_{\rm eff}\int_B d\Sigma_{\mu}g^{\mu\nu}\Psi_{\rm eff}^*\partial_{\nu}\Psi_{\rm eff}
    = i\int_B d\Sigma_{\nu}\nabla_{\mu}\mathcal{K}^{\mu\nu}\\
    = \frac{1}{4G\kappa}\int_{\partial B}d\Sigma_{\nu\mu}\nabla^{\nu}\xi^{\mu} = \frac{A}{4G}, 
\end{multline}
as required. The crucial result is that the geometric phase is given by $A/4G = 4\pi k$, as expected. Further considerations are considered beyond the scope of this work. 
\textit{\textbf{Entangled particles}} -- Since considering two polarization states for the gauge particles yields a factor of 2 (in $\nu(k) = 2k \in \mathbb{Z}$), by eq. (\ref{entropy_eq}), we conclude that the black hole is comprised of the two polarization degrees of freedom. However, near the horizon, an accelerating observer $\kappa = 1/4GM$ only has access to information on their Rindler wedge.\cite{cottrell2019build} This suggests we make use of $|k \rangle$ from  eq. (\ref{k_state_eq}) and consider thermal averages, 
\begin{subequations}\label{TFD_eq}
\begin{multline}
    \frac{1}{\sqrt{\mathcal{Z}}}\exp\left (\exp\left (-\frac{\pi\beta}{\lambda} \right )a_{2\pi/\lambda}^+a_{2\pi/\lambda}^- -\frac{\pi\beta}{2\lambda} \right )|k^+_{2\pi/\lambda}\rangle\otimes|k^-_{2\pi/\lambda}\rangle\\
    = \frac{1}{\sqrt{\mathcal{Z}}}\sum_{k = 0}^{\infty}\exp\left (-\frac{\pi\beta}{\lambda}(k + 1/2) \right )|k^+_{2\pi/\lambda}\rangle\otimes|k^-_{2\pi/\lambda}\rangle\\
    = |{\rm TFD} \rangle = \frac{1}{\sqrt{\mathcal{Z}}}\sum_{k = 0}^{\infty}\exp(-\beta E_k/2)|k \rangle,
\end{multline}
where $|{\rm TFD} \rangle$ is the Thermal Field Double state\cite{cottrell2019build},
\begin{align}
    E_k = \frac{2\pi}{\lambda}\left (k + \frac{1}{2}\right ),\\
    \mathcal{Z} = \sum_{k = 0}^{\infty}\exp\left (-\beta E_k \right ) = \frac{\exp(-\pi\beta/\lambda)}{1 - \exp(-2\pi\beta/\lambda)},
\end{align}
\end{subequations}
is the energy of the particles and the partition function respectively, and each particle polarization state is taken to occupy a given Rindler wedge (\textit{e.g.} ($+$) polarization occupies the right, $R$ while ($-$) polarization occupies the $L$ Rindler wedges) leading to quantum entanglement of the two left and right wedges. While the density matrix of the entangled gauge particles is a pure state ($\hat{\rho}^2 = 1$) given by, 
\begin{subequations}
\begin{align}\label{density_matrix_eq}
    \hat{\rho} = |{\rm TFD}\rangle \langle {\rm TFD}|,
\end{align}
the density matrix of each individual black hole is obtained by tracing out the quantum states of the other,
\begin{align}
    \hat{\rho}_{\pm} = \sum_{l_{\mp} = 0}^{\infty}\langle l_{\mp}|\hat{\rho}|l_{\mp} \rangle = \frac{1}{\mathcal{Z}}\sum_{k = 0}^{\infty}\exp \left ( -\beta E_k \right )|k_{\pm}\rangle\langle k_{\pm}|, 
\end{align}
\end{subequations}
which is a mixed/thermal state ($\hat{\rho}^2 \neq 1$). 

Since the black hole is constructed  purely by the gauge particles, it is reasonable to assume that the black hole information, or its thermal average $\langle \beta M \rangle /2$, can be calculated using $|{\rm TFD} \rangle$ in eq. (\ref{TFD_eq}). In other words, we wish to calculate the black hole entropy using eq. (\ref{TFD_eq}). Since the density matrix in eq. (\ref{density_matrix_eq}) is equivalent to,
\begin{multline}
    \hat{\rho} = | {\rm TFD}\rangle \langle {\rm TFD} |\\
    \equiv \rho_{kl} = \frac{1}{\mathcal{Z}}\exp\left ( -\frac{\pi\beta}{\lambda}(k + l + 1) \right ),
\end{multline}
one can check that it is a pure state by,
\begin{align}
    {\rm Tr}(\hat{\rho}) = \sum_{k,l = 0}^{\infty}\rho_{kl}\delta_{kl} = \sum_{k = 0}^{\infty} p_k = 1,\\
    {\rm Tr}(\hat{\rho}^2) = \sum_{k,l,m,n = 0}^{\infty}\rho_{km}\rho_{nl}\delta_{mn}\delta_{kl} = \sum_{k, l = 0}^{\infty} p_kp_l = 1,
\end{align}
where $\delta_{kl}$ is the Kronekar delta and 
\begin{align}
    p_k = \mathcal{Z}^{-1}\exp\left (-\frac{2\pi\beta}{\lambda}(k + 1/2)\right ),
\end{align}
the occupation probabilities. On the other hand, calculating the Von Newmann entropy yields,
\begin{multline}
    S_{\rm VN} = -{\rm Tr}(\hat{\rho}_{\pm}\ln(\hat{\rho}_{\pm})) = -\sum_{k = 0}^{\infty}p_k\ln p_k\\
    = \sum_{k,l = 0}^{\infty}\rho_{kl}\delta_{kl}\ln \mathcal{Z}  + \frac{\pi\beta}{\lambda}\sum_{k,l = 0}^{\infty}\rho_{kl}\delta_{kl}(k + l + 1)\\
    = \sum_{k = 0}^{\infty}p_k\ln \mathcal{Z}  + \frac{\pi\beta}{\lambda}\sum_{k = 0}^{\infty}p_k (2k + 1)\\
    = \ln \mathcal{Z} + \frac{2\pi}{g}\left (\langle k \rangle + \frac{1}{2}\right ),
\end{multline}
where we have used $g = \lambda/\beta$ from eq. (\ref{constant_eq}).

Recall that, $g > 0$ is the coupling constant, whose value until now has been considered real and positive ($\beta \geq 0$, $\lambda \geq 0$) but otherwise arbitrary. Since we are seeking the expression, $S_{\rm VN} \propto \langle k \rangle$, we ought to define $g$ such that the free energy, $\ln \mathcal{Z} = 0$ identically vanishes. Solving for $g$,
\begin{align}\label{vanish_Z_eq}
    \mathcal{Z} = \frac{\exp(-\pi/g)}{1 - \exp(-2\pi/g)} = 1,
\end{align}
yields the golden ratio, 
\begin{align}\label{golden_ratio_eq}
    \exp(\pi/g) = \frac{1 + \sqrt{5}}{2},
\end{align}
where we have discarded the negative solution which renders $g$ imaginary. Thus, we can rearrange the Von Newnann entropy to obtain,
\begin{multline}
    \langle \mathcal{I} \rangle = 4\pi \langle k \rangle = -2g\sum_{k = 0}^{\infty}p_k\ln p_k - 2\pi,\\
    = \frac{4\pi}{\exp(2\pi/g) - 1} = 4\pi\exp(-\pi/g).
\end{multline}
Crucially, the average information stored by the black hole on its surface is proportional to the entropy of the gauge particles, suggesting no information will be lost when the particles are emitted or absorbed by the black hole. Nonetheless, one needs to rigorously track the black hole formation and evaporation process for a definitive conclusion. 

\textit{\textbf{Dynamical space-time}} -- The adiabatic condition requires each instance in time to be approximated by the Schwarzschild solution which necessarily admits a time-like Killing vector but otherwise results in neglecting Hawking radiation entropy effects and hence unitarity. Ideas on extending our formalism begin by recalling that the relevant scales for quantum black holes in our approach are the Schwarzschild radius/inverse temperature/particle wavelength ($r_{\rm S} \sim \beta \sim \lambda$) and the inverse mass/Compton wavelength ($1/M \sim \lambda_{\rm C}$), where the adiabatic invariant/information content ($k(M, \lambda) \sim \mathcal{I}(M, \lambda)$) is the conserved quantity interpreted as the measure of black hole energy as earlier discussed. 

Consequently, standard calculus requires,
\begin{subequations}
\begin{align}
    d\mathcal{I}(M, \lambda) =  \frac{\partial \mathcal{I}(M, \lambda)}{\partial M}dM + \frac{\partial \mathcal{I}(M, \lambda)}{\partial \lambda}d\lambda,
\end{align}
where the system evolves exploring the $M, \lambda$ phase space. Heuristically, we can treat $M, \lambda$ as conjugate variables where we impose Hamilton's equations, 
\begin{align}
    \bar{m}_{\rm P}\frac{\partial \mathcal{I}}{\partial M} = \frac{d\lambda}{d\eta},\,\,\bar{m}_{\rm P}\frac{\partial \mathcal{I}}{\partial\lambda} = -\frac{dM}{d\eta},
\end{align}
\end{subequations}
with $\bar{m}_{\rm P}$ a constant with mass dimensions. This guarantees that $\mathcal{I}(M(\eta), \lambda(\eta))$ is a conserved quantity ($d\mathcal{I}(M(\eta), \lambda(\eta))/d\eta = 0$) over coordinate time, $\eta$. Thus, in analogy with classical Hamiltonian mechanics ($\partial H/\partial p = dx/dt, \partial H/\partial x = -dp/dt$, where $p, x$ are the conjugate momentum, position respectively), the relevant entropy becomes the black hole entropy, $4\pi GM^2$ less the entropy of Hawking radiation, $S_{\rm H.r.}(\lambda)$, 
\begin{align}\label{I_eq}
    \mathcal{I}(M, \lambda) = \frac{M^2}{2\bar{m}_{\rm P}^2} - S_{\rm H.r.}(\lambda),
\end{align}
where $\bar{m}_{\rm P} = 1/\sqrt{8\pi G}$ is the reduced Planck mass. Thus, since the radiation entropy must depend on particle energy, $\omega = 2\pi/\lambda$, the mass, $M$ is independent of time \textit{if and only if} there is no Hawking radiation ($dM/d\eta = \bar{m}_{\rm P}\partial S_{\rm H.r.}(\lambda)/\partial \lambda = 0$). 

For instance, the entropy of a quantum harmonic oscillator with two polarization states is given by, 
\begin{subequations}
\begin{align}
    S_{\rm h.o.}(\lambda) = 2\ln (\mathcal{Z}) + \frac{2\pi\beta}{\lambda}\coth(\pi\beta/\lambda),
\end{align}
where $\mathcal{Z} = (\exp(\pi\beta/\lambda) - \exp(-\pi\beta/\lambda))^{-1}$ is the partition function, $\beta$ the inverse temperature. Since the entropy only depends on $\lambda/\beta = g$ given in eq. (\ref{constant_eq}), it is unaffected by gravitational blue shift given in eq. (\ref{blue_shift_eq}). Moreover, for consistency with eq. (\ref{vanish_Z_eq}) in the appendix, we shall consider the special case, $\mathcal{Z} = 1$, which fixes the value of $g$ (see eq. (\ref{golden_ratio_eq})) and hence,
\begin{align}
    S_{\rm h.o.} = \frac{2\pi\beta}{\lambda}\coth(\pi\beta/\lambda) = S_{\rm VN}.
\end{align}
\end{subequations}
This discussion can be linked to the Friedmann-Lemaitre-Robertson-Walker (FLRW) metric that admits no time-like Killing vector, 
\begin{subequations}
\begin{align}
    d\tau^2 = -d\eta^2 + R^2(\eta)\left (\frac{dr^2}{1 \pm 8\pi K_0r^2} + r^2d\Omega^2\right ), 
\end{align}
where $d\Omega^2 = d\theta^2 + \sin^2\theta d\phi^2$ is the metric of the two-sphere, $R(\eta)$ is the scale factor and $K_0$ is a constant playing the role of Gaussian curvature. For Einstein Field Equations sourced by $T^{\mu\nu}$ (given in eq. (\ref{EFE_eq}) where $\rho \rightarrow 0$), the Friedmann equations correspond to eq. (\ref{energy_div_eq}) and,
\begin{align}
    \frac{1}{2}\left (\frac{\partial R}{\partial \eta} \right )^2 - \frac{G\left (\frac{4\pi}{3}R^3\langle T^{00} \rangle \right )}{R} = \pm 4\pi K_0,
\end{align}
\end{subequations}
which is interpreted as the analogue of eq. (\ref{I_eq}) when the space-time is dynamical by setting, $R(\eta) = \bar{m}_{\rm P}\lambda(\eta) = g\bar{m}_{\rm P}\beta(\eta)$, $\rho_0 = 2\pi\bar{m}_{\rm P}V^{-1}g\coth(\pi/g)$ and,
\begin{subequations}
\begin{align}
    \langle T^{00} \rangle = \frac{2\pi g}{\lambda VR^3}\coth(\pi\beta/\lambda) = \rho_0/R^4,\\
    V = \frac{4\pi\ell_{\rm P}^3}{3},\,\, S_{\rm H.r.} = \frac{S_{\rm VN}}{8\pi}\left (\frac{g}{R} \right )^2,
\end{align}
\end{subequations}
where $\ell_{\rm P} = 1/\bar{m}_{\rm P}$. In this case, the black hole information corresponds to, 
\begin{subequations}
\begin{align}
    \mathcal{I} = \pm 4\pi\ell_{\rm P}^2K_0,
\end{align}
interpreted as the analogue of the Gauss-Bonnet theorem, 
\begin{align}
    \mathcal{I} = \pm \int Kd(Area) = \pm 2\pi\chi(k) = 4\pi k,
\end{align}
\end{subequations}
where $\chi(h) = 2 - 2h(k)$ is the Euler characteristic of a two-dimensional compact orientable manifold of genus, 
\begin{align}\label{comm_eq}
    \pm (1 - h(k)) = k.
\end{align}

Moreover, since $k = a^{\dagger}a$ is the boson number, we ought to choose the negative sign corresponding to $h(k) = aa^{\dagger} \geq 1$ and eq. (\ref{comm_eq}) is merely the commutation relation,
\begin{subequations}
\begin{align}
    aa^{\dagger} - a^{\dagger}a = 1.
\end{align}
This implies that the positive sign corresponds to the case for fermions,
\begin{align}
    cc^{\dagger} + c^{\dagger}c = 1,
\end{align}
\end{subequations}
with $k = c^{\dagger}c$, $h(k) = cc^{\dagger} = 1 - k \leq 1$ and $T^{\mu\nu}$ replaced by the Dirac energy momentum tensor. Finally, the connection of eq. (\ref{I_eq}) to recent works on reproducing the Page curve using a generalized entropy equation for quantum extremal surfaces and bulk fields\cite{almheiri2019entropy, penington2020entanglement, engelhardt2021finding, engelhardt2015quantum} remains promising but nonetheless presently unexplored.

\end{document}